\documentclass{llncs}

\usepackage{amsmath}
\usepackage{graphicx}
\usepackage{color}

\begin{document}


\newcommand{\ket}[1]{| #1 \rangle}
\newcommand{\bra}[1]{\langle #1 |}
\newcommand{\braket}[2]{\langle #1 | #2 \rangle}

\newcommand{\comment}[1]{}

\newtheorem{Theorem}{Theorem}
\newtheorem{Lemma}{Lemma}
\newtheorem{Claim}{Claim}
\newtheorem{Corollary}{Corollary}
\newtheorem{Conjecture}{Conjecture}

\def\U{\Uparrow}
\def\D{\Downarrow}
\def\L{\Leftarrow}
\def\R{\Rightarrow}


\title{Quantum walks on two-dimensional grids with multiple marked locations}

\author{Nikolajs Nahimovs, Alexander Rivosh
\thanks{NN is supported by EU FP7 project QALGO, AR is supported by ERC project MQC.}
}
\institute{Faculty of Computing, University of Latvia, Raina bulv. 19, Riga, LV-1586, Latvia.} 

\maketitle


\textbf{Abstract}.

The running time of a quantum walk search algorithm depends on both the structure of the search space (graph) and the configuration of marked locations. While the first dependence have been studied in a number of papers, the second dependence remains mostly unstudied.

We study search by quantum walks on two-dimensional grid using the algorithm of Ambainis, Kempe and Rivosh [AKR05]. 
The original paper analyses one and two marked location cases only.
We move beyond two marked locations and study the behaviour of the algorithm for an arbitrary configuration of marked locations.

In this paper we prove two results showing the importance of how the marked locations are arranged. 
First, we present two placements of $k$ marked locations for which the number of steps of the algorithm differs by $\Omega(\sqrt{k})$ factor.
Second, we present two configurations of $k$ and $\sqrt{k}$ marked locations having the same number of steps and probability to find a marked location.


\section{Introduction}

Quantum walks are quantum counterparts of classical random walks \cite{Por13}. 
They have been useful to designing quantum algorithms for a variety of problems \cite{AKR05,MSS05,BS06,Amb07}. In many of those applications, quantum walks are used as a tool for search.

To solve a search problem using quantum walks, we introduce the notion of marked locations. Marked locations correspond to elements of the search space that we want to find. We then perform a quantum walk on the search space with one transition rule at unmarked locations and another transition rule at marked locations. If this process is set up properly, it leads to a quantum state in which marked locations have higher probability than the unmarked ones. This state can then be measured, finding a marked location with a sufficiently high probability. This method of search using quantum walks was first introduced in \cite{SKW03} and has been used many times since then.

The running time of a quantum walk search algorithm depends on both structure of the search space and the configuration --- the number and the placement --- of marked locations.
There have been a number of papers studying dependence of the running time on the structure of the graph.
Krovi \cite{Kro07} has studied symmetries of a graph and explained fast hitting times using symmetry concept. Janmark, Meyer and Wong \cite{JMW14} show that global symmetry of the graph is not necessary for fast quantum search. They demonstrate graphs with automorphism group consisting of an identity mapping only and still achieving $\Theta(\sqrt{N})$ quantum speed-up.
Mayer and Wong\cite{MW15} has studied connectivity of the graph and has shown that it is also a poor indicator of fast quantum search: there exists graphs with low connectivity but fast search, and graphs with high connectivity but slow search. So, despite of significant progress in the field the overall picture is still far from being complete. 

On the other hand, dependence on the number and the placement of marked locations remains mostly unstudied. Most of papers on quantum walk algorithms \cite{AKR05,APN14} prove their results for one or two marked locations only.

We study search by quantum walks on a finite two-dimensional grid using the algorithm of Ambainis, Kempe, Rivosh (AKR). The original \cite{AKR05} paper analyses the behaviour of the algorithm for one or two marked locations.
We move beyond two marked locations and study the behaviour of the algorithm for an arbitrary configuration of marked locations.
We show that the placement of marked locations has at least the same effect on the number of steps of the algorithm as the number of marked locations.

First, we present two placements of $k$ marked locations for which the number of steps of the algorithm differs by $\Omega(\sqrt{k})$ factor.
Here the first configuration is a block of $\sqrt{k} \times \sqrt{k}$ marked locations and the second configuration is $k$ uniformly distributed marked locations (placed at $\sqrt{N/k}$ distance from each other). 
We prove that the number of steps of the algorithm for the distributed placement is $\widetilde{O}(\sqrt{N/k})$, while for the grouped placement --- $\widetilde{\Omega}(\sqrt{N} - \sqrt{k})$.

Second, we present two configurations of $k$ and $\sqrt{k}$ marked locations, respectively, having the same number of steps and probability to find a marked location.
Here, the first configuration is a block of $\sqrt{k} \times \sqrt{k}$ marked locations and the second configuration is the perimeter of a $\sqrt{k} \times \sqrt{k}$ block (all internal locations are not marked).

Dependence of the number of steps on the placement of marked locations makes quantum walks different from Grover's search algorithm, where the number of steps have exact dependence on the number of marked locations. In case of quantum walks, even if the number of marked locations is known, the number of steps can vary depending on a placement of marked locations.
On the other hand, for all configurations studied in this paper, if the number of marked locations is in $[1,k]$ then the number of steps of the algorithm is still in $[\widetilde{O}(\sqrt{N/k}), \widetilde{O}(\sqrt{N})]$ --- same as it is for Grover's algorithm.


\section{Quantum walks in two dimensions}

Suppose we have $N$ items arranged on a two dimensional grid of size $\sqrt{N} \times \sqrt{N}$. We denote $n=\sqrt{N}$.
The locations on the grid are labelled by their $x$ and $y$ coordinate as $(x,y)$ for $x,y \in \{ 0, \dots, n-1\}$. We assume that the grid has periodic boundary conditions. For example, going right from a location $(n-1, y)$ on the right edge of the grid leads to the location $(0, y)$ on the left edge of the grid.

To introduce quantum version of random walk, we define a "location" register with basis states $\ket{i,j}$ for $i,j \in \{0,\dots,n-1\}$. Additionally, to allow non-trivial walks, we define a "direction" or "coin" register with four basis states, one for each direction: $\ket{\U}$, $\ket{\D}$, $\ket{\L}$ and $\ket{\R}$. Thus, the basis states of quantum walk are $\ket{i,j,d}$ for $i,j \in \{0,\dots,n-1\}$ and $d \in \{\U,\D,\L,\R\}$. The state of quantum walk is given by:
\\
$$
\ket{\psi(t)} = \sum_{i,j} (
\alpha_{i,j,\U}\ket{i,j,\U} + \alpha_{i,j,\D}\ket{i,j,\D} + 
\alpha_{i,j,\L}\ket{i,j,\L} + \alpha_{i,j,\R}\ket{i,j,\R} ).
$$

A step of the quantum walk is performed by first applying $I \times C$, where $C$ is unitary transform on the coin register. The most often used transformation on the coin register is the Grover's diffusion transformation $D$:

$$
D = \frac{1}{2} \left( 
\begin{array}{cccc}
-1 & 1 & 1 & 1 \\
1 & -1 & 1 & 1 \\
1 & 1 & -1 & 1 \\
1 & 1 & 1 & -1 
\end{array} \right).
$$
\\
Then, we apply the shift transformation $S$:
\\
$$
\begin{array}{lcl}
\ket{i,j,\U} & \rightarrow & \ket{i,j-1,\D} \\
\ket{i,j,\D} & \rightarrow & \ket{i,j+1,\U} \\
\ket{i,j,\L} & \rightarrow & \ket{i-1,j,\R} \\
\ket{i,j,\R} & \rightarrow & \ket{i+1,j,\L}
\end{array}
$$
\\
Notice that after moving to an adjacent location we change the value of the direction register to the opposite.

We start quantum walk in the state 
$$
\ket{\psi_0} = \frac{1}{\sqrt{4N}} \sum_{i,j} \big( \ket{i,j,\U} + \ket{i,j,\D} + \ket{i,j,\L} + \ket{i,j,\R} \big).
$$

It can be easily verified that the state of the walk stays unchanged, regardless of the number of steps. To use quantum walk as a tool for search, we "mark" some locations. For unmarked locations, we apply the same transformations as above. For marked locations, we apply $-I$ instead of $D$ as the coin flip transformation. The shift transformation remains the same in both marked and unmarked locations.

Another way to look at the step of the algorithm is that we first perform a query $Q$ transformation, which flips signs of amplitudes of marked locations, then conditionally perform the coin transformation -- $I$ or $D$ depending on whether the location is marked or not -- and then perform the shift transformation $S$.

If there are marked locations, the state of the algorithm starts to deviate from $\ket{\psi(0)}$. It has been shown \cite{AKR05} that after $O(\sqrt{N\log{N}})$ steps the inner product $\braket{\psi(t)}{\psi(0)}$ becomes close to $0$.

In case of one or two marked locations AKR algorithm finds a marked location with $O(1 / \log{N})$ probability. 
The probability is small, thus, the algorithm uses amplitude amplification to get $\Theta(1)$ probability. The amplitude amplification adds an additional $O(\sqrt{\log{N}})$ factor to the number of steps. Thus, the total running time of the algorithm is $O(\sqrt{N} \log{N})$.


\section{Results}


\subsection{Grouped and distributed placements of marked locations.}

In this subsection we show that the number of steps of the algorithm for two placements of $k$ marked locations can differs by $\Omega(\sqrt{k})$ factor.

\begin{figure}[ht]
\centering
\includegraphics[scale=0.5]{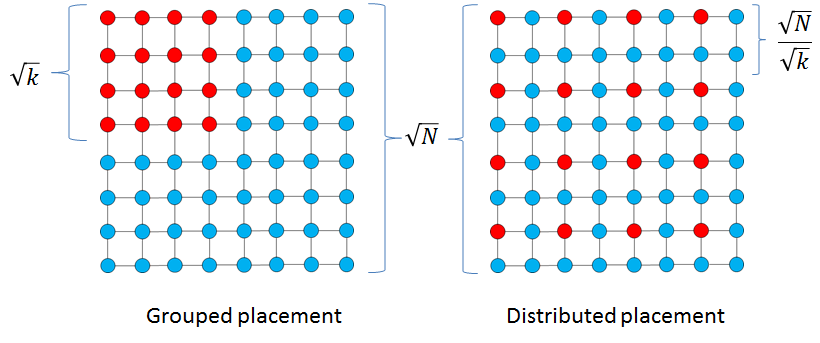}
\caption{Grouped and distributed placements of $k$ marked locations.}
\label{fig:grouped_vs_distributed}
\end{figure}

Consider two configurations (placements) of $k$ marked locations.
The first configuration is a block of $\sqrt{k} \times \sqrt{k}$ marked locations. The second configuration is $k$ uniformly distributed marked locations (placed at $\sqrt{N/k}$ distance from each other) (see figure \ref{fig:grouped_vs_distributed}). 
We will refer them as grouped and distributed placements respectively.

\begin{Lemma}[Distributed placement]
\label{lem:distributed_placement}
Let $k$ be a full square and let $k$ marked locations be uniformly distributed on $\sqrt{N} \times \sqrt{N}$ grid (placed at $\sqrt{N/k}$ distance from each other). Then AKR algorithm needs $O(\sqrt{N/k \cdot \log{(N/k)}})$ steps and finds a marked location with $O(1/\log{(N/k)})$ probability.
\end{Lemma}

\noindent
\textbf{Proof.}
By symmetry each of $\sqrt{N/k} \times \sqrt{N/k}$ regions of the grid is experiencing the same evolution (here by region we mean a part of the grid with a marked location in its top-left corner).

More formally,
consider basis states corresponding to locations on $\sqrt{N/k}$ distance from each other pointing to the same direction.
Initially amplitudes of all such pairs of basis states are equal.
For each pair of basis states the step of the algorithm applies the same transformations to the same amplitudes. Thus, after the step of the algorithm amplitudes of a pair of basis states are also equal.
Therefore, the evolution of each of $\sqrt{N/k} \times \sqrt{N/k}$ regions of the grid is essentially the same.

We have $k$ copies of quantum walk on $\sqrt{N/k} \times \sqrt{N/k}$ grid with a single marked location.
Therefore, after
$O(\sqrt{N/k \cdot \log{(N/k)}})$ steps --- the number of steps for the  $\sqrt{N/k} \times \sqrt{N/k}$ grid with a single marked location --- overlap of the current and the initial states of the algorithm becomes close to $0$. If we measure the state at this point the probability to get one of basis states corresponding to a marked location is $O(1/\log{(N/k)})$.
\qed

\begin{Lemma}[Grouped placement]
\label{lem:grouped_placement}
Let $k$ be a full square and let $k$ marked locations be placed as a $\sqrt{k} \times \sqrt{k}$ square on $\sqrt{N} \times \sqrt{N}$ grid. Then AKR algorithm needs $\Omega(\sqrt{N} - \sqrt{k})$ steps.
\end{Lemma}

\noindent
\textbf{Proof.}
This follows from the fact that the average distance from a location on the grid to a marked location is $\Omega(\sqrt{N} - \sqrt{k})$. Thus, the algorithm needs at least this number of steps to achieve a constant probability of finding a marked location.
\qed

We have shown that the number of steps for the grouped and the distributed placements differ by $\Omega(\sqrt{k})$ factor.
The grouped and the distributed placements are two extreme cases, therefore, we believe that $O(\sqrt{k})$ is the maximal possible gap for any two placements of $k$ marked locations. We conjecture

\begin{Conjecture}
Let $P_1$ and $P_2$ be two placements of $k$ marked locations on $\sqrt{N} \times \sqrt{N}$ grid. Then the number of steps of AKR algorithm for $P_1$ and $P_2$ can differ by at most $O(\sqrt{k})$ factor.
\end{Conjecture}


\subsection{Evolution of amplitudes of near-by marked locations.}

In the previous subsection we showed that AKR algorithm is inefficient for grouped marked locations. The reason for this is the coin transformation, which does not rearrange amplitudes within a marked location. Therefore, marked locations inside the group have almost no effect on the number of steps and the probability to find a marked location of the algorithm.

In this subsection we explore grouped marked locations in more details. We analyse the evolution of amplitudes of two near-by marked locations. We show, that the step of AKR algorithm do not change absolute values of adjoint amplitudes of near-by marked locations.

\begin{Theorem}
\label{thm:near_by_marked_locations}
Let $\ket{\psi(t)}$ be a state of AKR algorithm after $t$ steps and let locations $(i,j)$ and $(i,j+1)$ be marked. Then for any $t$ we have $$\braket{\psi(t)}{i,j,\R} = \braket{\psi(t)}{i,j+1,\L} = (-1)^t/\sqrt{4N}.$$
\end{Theorem}

\begin{figure}[ht]
\centering
\includegraphics[scale=0.5]{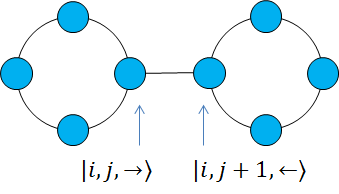}
\caption{Amplitudes of near-by marked locations.}
\label{fig:near_by_marked_locations}
\end{figure}

\noindent
\textbf{Proof.}
Consider the effect of the step of the algorithm on amplitudes of $\ket{i,j,\R}$ and $\ket{i,j+1,\L}$. Query changes signs of both amplitudes (both locations are marked); coin flip does nothing (both locations are marked); shift swaps the amplitudes. More formally,

$$
\begin{array}{lll}
Q\ket{i,j,\R} = -\ket{i,j,\R} & \quad & 
Q\ket{i,j+1,\L} = -\ket{i,j+1,\L} \\
C\ket{i,j,\R}= \ket{i,j,\R} & \quad & 
C\ket{i,j+1,\L} = \ket{i,j+1,\L} \\
S\ket{i,j,\R}= \ket{i,j+1,\L} & \quad & 
S\ket{i,j+1,\L} = \ket{i,j,\R}
.
\end{array}
$$
Therefore, the step of the algorithm changes signs of the amplitudes and swaps their values.

Initially all amplitudes are equal to $1/\sqrt{4N}$. Thus, the values of the amplitudes will be $1/\sqrt{4N}$ after odd number steps and $-1/\sqrt{4N}$ after even number steps. 
\qed


\subsection{Block and perimeter configurations of marked locations.}

In this subsection we present two configurations of $k$ and $\sqrt{k}$ marked locations, respectively, having the same number of steps and probability to find a marked location.

\begin{figure}[ht]
\centering
\includegraphics[scale=0.5]{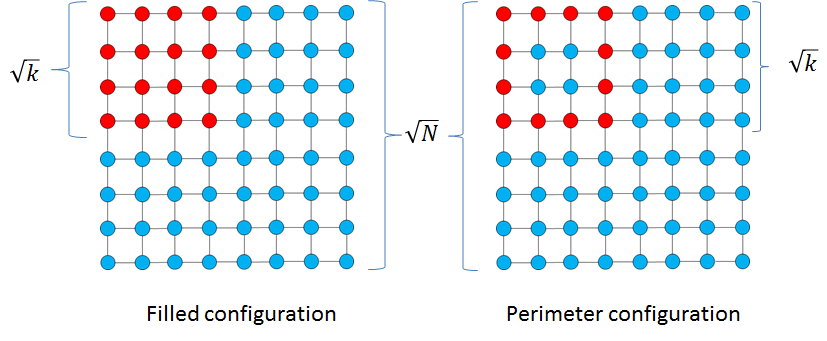}
\caption{Filled and perimeter configurations of marked locations.}
\label{fig:filled_vs_perimeter}
\end{figure}

Consider two configurations of marked locations: $k$ marked locations placed as a $\sqrt{k} \times \sqrt{k}$ square and $4(\sqrt{k}-1)$ marked locations placed as the perimeter of a $\sqrt{k} \times \sqrt{k}$ square (figure \ref{fig:filled_vs_perimeter}).
We will refer them as filled and perimeter configuration respectively.

\begin{figure}[ht]
\centering
\includegraphics[scale=0.4]{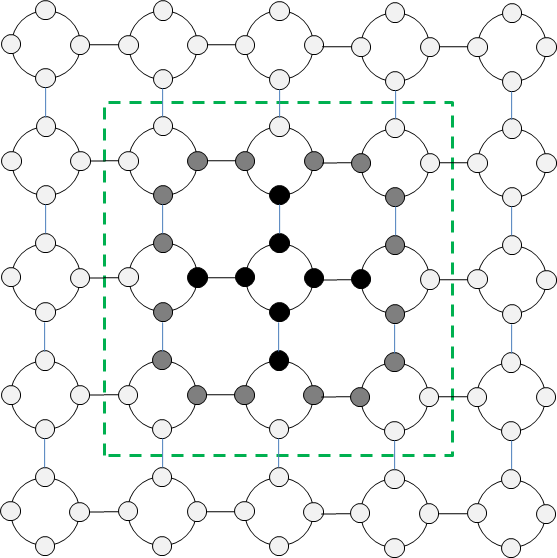}
\caption{Group of $3 \times 3$ marked locations (in the center). Basis states of $\ket{\psi_{out}(t)}$ are coloured with light gray, basis states of $\ket{\psi_{per}(t)}$ are coloured with dark gray, basis states of $\ket{\psi_{in}(t)}$ are coloured with black.}
\label{fig:parts_of_the_state}
\end{figure}

Let $\ket{\psi(t)}$ be the state of AKR algorithm after $t$ steps for the filled configuration and $\ket{\phi(t)}$ be the state of AKR algorithm after $t$ steps for the perimeter configuration.
For the further analysis we split $\ket{\phi(t)}$ into three parts (figure \ref{fig:parts_of_the_state}):
 
\begin{itemize}
\item
$\ket{\psi_{out}(t)}$ consisting of basis states of the outer part of the square as well as basis states of the perimeter pointing to the outer part
\item 
$\ket{\psi_{in}(t)}$ consisting of basis states of the inner part of the square as well as basis states of the square pointing to the inner part
\item
$\ket{\psi_{per}(t)}$ consisting of basis states of the perimeter pointing to other locations on the perimeter. 
\end{itemize}
Similarly we define $\ket{\phi_{out}(t)}$, $\ket{\phi_{in}(t)}$ and $\ket{\phi_{per}(t)}$.

\begin{Lemma}
$\forall t \geq 0: \ket{\psi_{per}(t)} = \ket{\phi_{per}(t)}$. 
\end{Lemma}

\noindent
\textbf{Proof.}
According to theorem \ref{thm:near_by_marked_locations} all amplitudes of basis states of $\ket{\psi_{per}(t)}$ and $\ket{\phi_{per}(t)}$ are equal to $(-1)^t/\sqrt{4N}$.
\qed

\begin{Lemma}
$\forall t \geq 0: \ket{\psi_{out}(t)} = \ket{\phi_{out}(t)}$.
\end{Lemma}

\noindent
\textbf{Proof.} 
Consider the effect of a step of the algorithm on $\ket{\psi(t)}$.
First, consider coin transformation.
For the outer (the inner) part of the square it acts on the basis states belonging $\ket{\psi_{out}(t)}$ ($\ket{\psi_{in}(t)}$) only. For the perimeter of the square it acts on all three parts. However, as the locations on the perimeter are marked and coin transformation for marked locations is equal to $-I$, amplitudes of basis states belonging to different parts do not interact with each other. 
Next, consider shift transformation.
For each of parts of the state shift swaps amplitudes within the part.
Therefore, the step of the algorithm acts on each part of the state independently of other parts.
In other words, evolution of each part of $\ket{\psi(t)}$ is independent on evolutions of other parts.
The above argument holds for $\ket{\phi(t)}$ without any changes.

Initially $\ket{\psi_{out}(0)} = \ket{\phi_{out}(0)}$. There is no marked locations in the outer part of the square. Thus, the transformation applied to $\ket{\psi_{out}(t)}$ and $\ket{\phi_{out}(t)}$ are the same.
Therefore, $\ket{\psi_{out}(t)} = \ket{\phi_{out}(t)}$ will hold for all $t$.
\qed

Next theorem estimates the overlap between the state of the algorithm after $t$ steps for the filled and the perimeter configurations.

\begin{Theorem}
\label{thm:filled_vs_perimeter_overlap}
$ \forall t \geq 0: \braket{\psi(t)}{\phi(t)} \geq 1 - \Theta(k/N)$.
\end{Theorem}

\noindent
\textbf{Proof.}
$$ 
\braket{\psi(t)}{\phi(t)} = 
\braket{\psi_{out}(t)}{\phi_{out}(t)} + 
\braket{\psi_{in}(t)}{\phi_{in}(t)} + 
\braket{\psi_{per}(t)}{\phi_{per}(t)}
.
$$ 
It follows from the previously proved lemmas that the only parts of $\ket{\psi(t)}$ and $\ket{\phi(t)}$ which may differ are $\ket{\psi_{in}(t)}$ and $\ket{\phi_{in}(t)}$. Thus,
$$ 
\braket{\psi(t)}{\phi(t)} = 1 - 
\braket{\psi_{in}(t)}{\psi_{in}(t)} + 
\braket{\psi_{in}(t)}{\phi_{in}(t)}
.
$$ 
Amplitudes of basis states of $\ket{\psi_{in}(t)}$ are equal to $(-1)^t/\sqrt{4N}$.
There are $(\sqrt{k}-2)^2$ inner locations with four amplitudes each and $4(\sqrt{k}-2)$ amplitudes of the perimeter pointing to inner locations.
The total number of basis states in $\ket{\psi_{in}}$ is
$$ c(k) = 4(\sqrt{k}-2)^2 + 4(\sqrt{k}-2) = 4(k -3\sqrt{k} + 2) $$ 
and, thus, we have
$$ 
\braket{\psi(t)}{\phi(t)} =
1 - \frac{c(k)}{4N} + \braket{\psi_{in}(t)}{\phi_{in}(t)}.
$$
$\braket{\psi_{in}(t)}{\phi_{in}(t)}$ can take values from $[ -\frac{c(k)}{4N}, \frac{c(k)}{4N} ]$. Therefore,
$$ 
\braket{\psi(t)}{\phi(t)} \geq 
1 - 2 \cdot \frac{c(k)}{4N} = 1 - \Theta\left(\frac{k}{N}\right).
$$
\qed

Now we give a corollary of the above theorem which bounds the maximal difference in the number of steps of the algorithm for the configurations.
Note, that we are interested in the case $k = o(N)$. Otherwise, if $k$ is of the same order as $N$ then the trivial ``measure on the first step'' approach finds a marked location with constant probability.

\begin{Corollary}
\label{cor:filled_vs_perimeter_steps}
Let $k = o(N)$. Then the number of steps of AKR algorithm for the filled and the perimeter placements can differ by at most one.
\end{Corollary}

\noindent
\textbf{Proof.}
It follows from \cite{Sze04} that the number of steps of AKR algorithm can not increase if we mark a previously unmarked location. Therefore, the total number of steps for $k$ marked locations is at most $O(\sqrt{N \log{N}})$ --- the number of steps of the algorithm for a single marked location. 
The angle between the state for the filled and the perimeter configurations is less than the angle to which the state is rotated by the step of the algorithm. Thus, the number of steps of the algorithm for the configurations can differ by at most one.
\qed

Next theorem estimates the maximal difference in the probability to find a marked location after $t$ steps for the filled and the perimeter configurations.

\begin{Theorem}
\label{thm:filled_vs_perimeter_probability}
$\forall t \geq 0$: probability to find a marked location for $\ket{\psi(t)}$ and $\ket{\phi(t)}$ differs by at most $\Theta(k/N)$.
\end{Theorem}

\noindent
\textbf{Proof.}
It follows from the previously proved lemmas that the only parts of $\ket{\psi(t)}$ and $\ket{\phi(t)}$ which may differ are $\ket{\psi_{in}(t)}$ and $\ket{\phi_{in}(t)}$. 
For the filled configuration all amplitudes of $\ket{\psi_{in}(t)}$ are equal to $(-1)^t/\sqrt{4N}$. For the perimeter configuration inner part is not marked. Additionally, amplitudes of the perimeter pointing to the inner part might become zero. Thus, the maximal possible difference in probability to measure a marked location is
$ \frac{1}{4N} \cdot c(k) = \Theta\left(\frac{k}{N}\right) $.
\qed

A typical probability for AKR algorithm to find a marked location is $\Omega(1/\log{N})$. Thus, the probability to find a marked location for the configurations differs by an insignificant factor.

We have shown that for the filled and the perimeter configurations of marked locations AKR algorithm has the same number of steps and the  probability to find a marked locations. However, the filled configuration has quadratically larger number of marked locations than the perimeter configuration.


\section{Conclusions and discussion}

In this paper we analysed AKR quantum walk search algorithm for two-dimensional grid with multiple marked locations.
First, we showed that the placement of $k$ marked locations can change the number of steps of the algorithm by $\Omega(\sqrt{k})$ factor.
Namely, we showed that the number of steps of the algorithm for the grouped placement ($k$ marked locations are placed as $\sqrt{k} \times \sqrt{k}$ group) is $\widetilde{O}(\sqrt{N/k})$, while for the distributed placement (marked locations are placed at $\sqrt{N/k}$ distance from each other) is $\widetilde{\Omega}(\sqrt{N} - \sqrt{k})$.

The proved result shows that the number of steps for $k$ marked locations can be in range $[\widetilde{O}(\sqrt{N/k}), \widetilde{\Omega}(\sqrt{N})]$.
We conjecture that this is the maximal possible gap and the number of steps of the AKR algorithm for two placements of $k$ marked locations can differ by at most $O(\sqrt{k})$.

It would be interesting to extend the analysis to three and more-dimensional grids, as while our argument for the distributed placement  still holds for higher dimensions, the argument for the grouped placement is bound to two-dimensional case.

Second, we present two configurations of $k$ and $\sqrt{k}$ marked locations, respectively, having the same number of steps and probability to find a marked location.
Here, the first configuration is a block of $\sqrt{k} \times \sqrt{k}$ marked locations and the second configuration is the perimeter of a $\sqrt{k} \times \sqrt{k}$ block (all internal locations are not marked).
We showed that marked locations inside the block have almost no effect on the number of steps and the probability to find a marked location of the algorithm. More formally, we showed that internal locations of the block do not contribute to the growth of probability to find a marked location as well as do not affect the number of steps of the algorithm. Thus, the proved result holds not just for square blocks, but for any block of marked locations.
Our analysis includes a number of support theorems and observations that might be of independent interest.



\end{document}